\documentstyle[12pt,fleqn]{article}
\input epsf
\def\ktilde{{{\tilde{k}}}}
\def\etatilde{{{\tilde{\eta}}}}
\def\be{\begin{equation}}
\def\ee{\end{equation}}
\def\ba{\begin{eqnarray}}
\def\ea{\end{eqnarray}}
\def\fun#1#2{\lower3.6pt\vbox{\baselineskip0pt\lineskip.9pt
        \ialign{$\mathsurround=0pt#1\hfill##\hfil$\crcr#2\crcr\sim\crcr}}}
\def\la{\mathrel{\mathpalette\fun <}}
\def\ga{\mathrel{\mathpalette\fun >}}
\def\mpl{{{M_{{\rm Pl}}}}}
\def\eqr#1{{Eq.\ (\ref{#1})}}

\begin{document}

\begin{titlepage}
\vspace*{-64pt}
\begin{flushright} {\footnotesize
FERMILAB-Pub-98/021--A\\
CERN-TH/98-37\\
OUTP-98-02-P\\
hep-ph/9802238 \\ }
\end{flushright}

\vskip 1.5cm

\begin{center}
{\Large\bf  Superheavy dark matter \\}

\vskip 1cm
{\bf 
Daniel J. H. Chung,$^{a,b,}$\footnote{E-mail: {\tt djchung@yukawa.uchicago.edu}}
 Edward W. Kolb,$^{b,c,}$\footnote{E-mail: {\tt  rocky@rigoletto.fnal.gov}}
 Antonio Riotto$^{d,}$\footnote{E-mail: {\tt riotto@nxth04.cern.ch}}$^,$\footnote{
     On leave  from Department of Theoretical Physics,
     University of Oxford, U.K. }
}
\vskip .75cm
{\it 
$^a$Department of Physics and Enrico Fermi Institute \\ 
The University of Chicago, Chicago, Illinois 60637-1433\\
\vspace{12pt}
$^b$NASA/Fermilab Astrophysics Center \\ Fermilab
National Accelerator Laboratory, Batavia, Illinois~~60510-0500\\
\vspace{12pt}
$^c$Department of Astronomy and Astrophysics and  Enrico Fermi Institute\\
The University of Chicago, Chicago, Illinois~~60637-1433\\
\vspace{12pt}
$^d$Theory Division, CERN, CH-1211 Geneva 23, Switzerland
}
\end{center}
\vskip .5cm

\begin{quote}
%We show that, contrary to the standard lore, dark matter may be
%superheavy (many orders of magnitude larger than the weak scale).  We
%show that massive particles may be produced naturally during the
%transition from the inflationary phase to either a matter-dominated or
%radiation-dominated phase as a result of the expansion of the
%background spacetime acting on vacuum quantum fluctuations of the dark
%matter field.  We find that as long as there are stable particles
%whose mass is of the order of the inflaton mass (presumably around
%$10^{13}$GeV), they will be produced in sufficient abundance to give
%$\Omega_0 =1$ quite independently of any details of the
%non-gravitational interactions of the dark-matter field.

We show that in large-field inflationary scenarios, superheavy (many
orders of magnitude larger than the weak scale) dark matter will be
produced in cosmologically interesting quantities if superheavy stable
particles exist in the mass spectrum.  We show that these particles
may be produced naturally during the transition from the inflationary
phase to either a matter-dominated or radiation-dominated phase as a
result of the expansion of the background spacetime acting on vacuum
quantum fluctuations of the dark matter field.  We find that as long
as there are stable particles whose mass is of the order of the
inflaton mass (presumably around $10^{13}$GeV), they will be produced
in sufficient abundance to give $\Omega_0 =1$ quite independently of
any details of the non-gravitational interactions of the dark-matter
field.
\end{quote}
\end{titlepage}

\baselineskip=24pt

%\renewcommand{\baselinestretch}{1.5}

%%%%%%%%%%%%%%%%%%%%%%%%%%%%%%%%%%%%%%%%
%%%%%%%%%%%%%%%%%%%%%%%%%%%%%%%%%%%%%%%%
\vspace{36pt}
\centerline{\bf I. INTRODUCTION}
\vspace{24pt}
%%%%%%%%%%%%%%%%%%%%%%%%%%%%%%%%%%%%%%%%
%%%%%%%%%%%%%%%%%%%%%%%%%%%%%%%%%%%%%%%%

It is now commonly accepted that most of the mass in galactic
halos as well as in the Universe as a whole is composed of dark matter
(DM). There are many indications that the DM consists of some new, and
yet undiscovered, weakly interacting massive particles (WIMPs).

Despite the fact that the nature of the DM is still unknown, it is
usually thought that DM particles cannot be too heavy.  If the WIMP is
a thermal relic, then it was once in local thermodynamic equilibrium
(LTE) in the early universe, and its present abundance is determined
by its self-annihilation cross section.  From unitarity arguments
\cite{griestkam}, one expects the mass of a thermal relic to be less
than about 500 TeV.  The present abundance of non-thermal relics is
not determined by their self-annihilation cross section since they
needn't have been ever in LTE in the early universe.  An example of a
non-thermal relic is the axion, and the present axion abundance is
determined by the dynamics of the phase transition associated with
symmetry breaking.  Non-thermal relics are typically very light, e.g.,
the axion mass is expected to be in the range $ 10^{-5}$ to $10^{-2}$eV
\cite{raf}.

Because the assumption of relatively low-mass DM seems quite natural,
it is rarely questioned.\footnote{Of course, superheavy dark matter
particles have been considered before to a certain extent.  In
particular, there is an extensive literature regarding observational
constraints on unusually heavy dark matter candidates (for example,
see Refs.\ \cite{ellis}, \cite{sarkar}, \cite{rujula}, and references
therein).  However, they do not restrict our scenario nor do they
consider our production mechanism.} The goal of this paper is to show
that the Universe might be made of superheavy WIMPs (we will refer to
them as $X$ particles), with mass larger than the weak scale by
several (perhaps many) orders of magnitude. Two conditions are
necessary for this to happen: {\it a)} the $X$ particles must be
cosmologically stable and {\it b)} their interaction rate must be
sufficiently weak such that thermal equilibrium with the primordial
plasma was never obtained.  This second condition is easy to satisfy
as long as the particle is extremely massive (of the order of the
Hubble parameter at the end of inflation).

We point out that superheavy dark matter may be created during the
evolution of the Universe in a number of ways.  If it is produced
during the process of reheating after inflation, then the upper bound
on its mass $M_X$ can be as large as the reheating temperature
$T_{RH}$. The latter should be less than about $10^9$GeV in order to
avoid overproducing dangerous relics such as quasistable gravitinos in
supergravity inspired scenarios.  The mass upper bound can be pushed
higher than the reheating temperature if one allows the DM to be
produced directly through the decay of the inflaton field.  In that
case, the mass upper bound is the inflaton field mass, which is
presumably less than about $10^{13}$GeV.  On the other hand, if
reheating after inflation is preceded by a preheating stage
\cite{preheating} it is certainly possible to produce by resonance
effects copious amounts of dark matter particles with masses much
larger than the inflaton mass
\cite{heavy}.

In this paper, we consider yet another mechanism of generating heavy
DM.  We study the possibility that DM is produced in the transition
between an inflationary and a matter-dominated (or
radiation-dominated) universe due to the ``nonadiabatic'' expansion of
the background spacetime during the transition acting on the vacuum
quantum fluctuations.

The distinguishing feature of this mechanism is the capability of
generating particles with mass of the order of the inflaton mass
(usually much larger than the reheating temperature) even when the
particles only interact extremely weakly (or not at all) with other
particles and do not couple to the inflaton(s).  We find that they may
still be produced in sufficient abundance to achieve critical density
today due to the classical gravitational effect on the vacuum state at
the end of inflation.  More specifically, we will show that in the
range $0.04 \la M_X/H \la 2$, where $H \sim m_\phi \sim 10^{13}$GeV is
the Hubble constant at the end of inflation ($m_\phi$ being the mass
of the inflaton), the DM produced gravitationally can have a density
today of the order of the critical density. This result is quite
robust with respect to the ``fine'' details of the transition between
the inflationary phase and the matter-dominated phase, and independent
of the coupling of the DM to any other particle.  This result is
reasonably robust also with respect to the ambiguity associated with
the choice of the vacua as we have tried to minimize the number of
particles produced by choosing an infinite adiabatic order in-out
vacua.  The only ``non-trivial'' requirements, other than that large
field inflation occurs, are that the WIMPs posses a mass close to the
inflaton mass and that they are stable.

Mechanically, the DM particle creation scenario is similar to the
inflationary generation of gravitational perturbations that seed the
formation of large scale structures (see for example the review given
in Ref.\ \cite{mukhanov}).  In the usual scenarios of this form,
however, the quantum generation of energy density fluctuations from
inflation is associated with the inflaton field which dominated the
mass density of the universe, and not a generic, sub-dominant scalar
field.

Because it is usually assumed that DM forms from the decays or
interactions of the reheating products, it usually has a stage of LTE
in its early history.  In our scenario the large mass of the
dark-matter particle will prevent it from thermalizing, and its
abundance will depend only on its mass and the behavior of the
spacetime, not on its weak coupling to other nongravitational fields.

Others have considered gravitational particle production at the end of
inflation.  For example, Ford \cite{ford} and Yajnik \cite{yajnik}
both consider particle production as a result of the nonadiabaticity
of the transition from an inflationary phase to a matter or radiation
dominated phase (although with a different cosmological implication in
mind).  Ford treats only massless, non-conformally coupled fields
using a well known perturbation technique (see references within
\cite{ford}).  Yajnik considers minimally coupled scalar field theory
in the limit of small masses, with an abrupt transition from an
inflationary phase to a radiation dominated phase.  In our work, we
consider extremely massive, conformally coupled fields and calculate
the particle production exactly by numerically solving the mode
equation.  We treat the conformally coupled case because conformal
coupling generally minimizes the number of particles produced,
particularly in the small mass ranges.  Unlike Yajnik, we also
consider the case where the metric is an analytic function of the
conformal time and show that this leads to qualitatively different
behavior of the density of particles produced for large masses.  The
analyticity implies a conservative estimate since fewer particles are
produced in that case than in the abrupt transition case.

Some of the ideas present in our scenario are also contained in the
work of Linde and Kofman \cite{linde1985}, \cite{kofman1986}, and
\cite{kofmanlinde1987}.  However, the purpose of their work was
to point out that isocurvature cosmological (large scale)
perturbations can be produced during inflation.  They did not consider
the importance of the nonadiabaticity of the transition at the end of
inflation which is responsible for the production of our superheavy
dark matter.  Instead, they mainly relied upon estimates of the particle
production during the de Sitter phase or the classical (long
wavelength) component of the particle field energy density left over
after inflation.

This paper is organized as follows.  In the next section, we elaborate
on the dark matter scenario and the calculational method.  In Section
III, we discuss the numerical results.  We then summarize our work in
Section IV.  In the appendix, we derive the asymptotic mass dependence
of the dark matter density presented in Section II.

%\newpage

%%%%%%%%%%%%%%%%%%%%%%%%%%%%%%%%%%%%%%%%
%%%%%%%%%%%%%%%%%%%%%%%%%%%%%%%%%%%%%%%%
\vspace{36pt}
\centerline{\bf II. SCENARIO AND CALCULATIONAL METHOD}
\vspace{24pt}
%%%%%%%%%%%%%%%%%%%%%%%%%%%%%%%%%%%%%%%%
%%%%%%%%%%%%%%%%%%%%%%%%%%%%%%%%%%%%%%%%

In this section we discuss the dark matter abundance calculation in
our scenario.  First, we give an expression for the dark matter
density today in terms of the number density when it was produced.  We
then consider the mass range of the dark matter necessary if it is
never to thermalize.  Finally, we discuss the mechanics of the
gravitational production of particles.  In particular, we discuss the
number density definition and present the asymptotic dependence of the
number density on the particle mass.

Suppose the dark matter never attains LTE and is nonrelativistic at
the time of production.  The usual quantity $\Omega_X h^2$ associated
with the dark matter density today can be related to the dark matter
density when it was produced.  To develop the relation, we begin by
writing
\begin{equation} 
\frac{\rho_X(t_0)}{
\rho_R(t_0)}
=\frac{\rho_X(t_{RH})}{\rho_R(t_{RH})}\:\left(\frac{T_{RH}}{T_0}\right),
\label{eq:transfromrh}
\end{equation}
where $\rho_R$ denotes the energy density stored in radiation,
$\rho_X$ denotes the energy density residing in the dark matter,
$T_{RH}$ is the reheating temperature, $T_0$ is the temperature today,
$t_0$ denotes the time today, and $t_{RH}$ denotes the approximate
time of reheating completion.\footnote{More specifically, this is
approximately the time at which the Universe becomes radiation
dominated.} To obtain $\rho_X(t_{RH})/\rho_R(t_{RH})$, we must
determine when $X$ particles are produced with respect to the
completion of reheating and the effective equation of state operative
between $X$ production and the completion of reheating.

At the end of inflation the universe may have a brief period of matter
domination resulting either from the coherent oscillations phase of
the inflaton condensate or from the preheating phase
\cite{preheating}.  If the $X$ particles are produced at time
$t=t_{e}$ when the de Sitter phase ends and the coherent oscillation
period just begins, then both the $X$ particle energy density and the
inflaton energy density will redshift at approximately the same rate
until reheating is completed and radiation domination begins.  Hence,
the ratio of energy densities preserved in this way until the time of
radiation domination is
\begin{equation} 
\frac{\rho_X(t_{RH})}{ \rho_R(t_{RH})} \approx \frac{8\pi}{3}\:
\frac{\rho_X(t_{e})}{\mpl^2 H^2(t_{e}) },
\end{equation} 
where $\mpl \approx 10^{19}$ GeV is the Planck mass and most of the
energy density in the universe just before time $t_{RH}$ is presumed
to turn into radiation.  Thus, using \eqr{eq:transfromrh}, we may get
an expression for the quantity $\Omega_X\equiv
\rho_X(t_0)/\rho_C(t_0)$, where $\rho_C(t_0)=3 H_0^2\mpl^2/8\pi$ and
$H_0=100\: h$ km sec$^{-1}$ Mpc$^{-1}$:
\begin{equation} 
\Omega_X h^2 \approx \Omega_R h^2\:
\left(\frac{T_{RH}}{T_0}\right)\: 
\frac{8 \pi}{3} \left(\frac{M_X}{\mpl}\right)\:
\frac{n_X(t_{e})}{\mpl H^2(t_{e})}.
\label{eq:omegachi}
\end{equation}
Here $\Omega_R h^2 \approx 4.31 \times 10^{-5}$ is the fraction of
critical energy density that is in radiation today and $n_X$ is the
density of $X$ particles at the time when they were produced.

Note that because the reheating temperature must be much greater than
the temperature today ($T_{RH}/ T_0 \ga 4.2 \times 10^{14}$), in order
to satisfy the cosmological bound $\Omega_X h^2 \la 1$, the fraction
of total energy density in the dark matter at the time when they were
produced must be extremely small.  To illustrate this, take
$H^2(t_{e}) \sim m_\phi^2$ and $\rho(t_e) \sim m_\phi^2\mpl^2$.  Then 
$\Omega_Xh^2 \sim 10^{17}(T_{RH}/10^9\mbox {GeV})(\rho_X(t_e)/\rho(t_e))$.  
It is indeed a very small fraction of the total energy density we wish to 
extract in the form of massive $X$ particles.

This means that if the dark matter particle is extremely massive, the
challenge lies in creating very few of them naturally.  We will see
that the gravitational production naturally gives the needed
suppression.  Note that if reheating occurs abruptly at the end of
inflation, then the matter domination phase may be negligibly short
and the radiation domination phase may follow immediately after the
end of inflation.  However, this does not change \eqr{eq:omegachi}.

For the superheavy $X$ particles to be good candidates for DM, they
have to be stable or at least have a lifetime greater than the age of
the universe.  This may occur in supersymmetric theories where the
breaking of supersymmetry is communicated to the ordinary sparticles
via the usual gauge forces \cite{reviewdsb}. In gauge-mediated
supersymmetric models there are two sectors with possible stable
particles which might act as superheavy dark matter candidates:

1) The secluded sector, which is strongly interacting: Supersymmetry
is broken dynamically and some $F$-term gets a nonvanishing
expectation value, where the scale of supersymmetry breaking, as
usual, is denoted by $\sqrt{F}$.

2) The messenger sector: This sector contains the fields charged under
the $SU(3)_C\otimes SU(2)_L\otimes U(1)_Y$ gauge interactions, and
communicate supersymmetry breaking to the sparticles in the observable
sector. The mass of the messenger fields is usually denoted by $M$.

After the messengers have been integrated out, sfermions receive a
mass squared $\widetilde{m}^2 \sim \alpha^2 \Lambda^2$, where $\alpha$
is the appropriate gauge coupling and $\Lambda\simeq F/M$. Notice, in
particular, that the spectrum of the superparticles depends on the
ratio $\Lambda= F/M$ which is fixed to be relatively small and in the
range $10$ to $10^3$ TeV. However, this does not necessarily mean that
$\sqrt{ F}$ and $M$ are of the same order of magnitude as $\Lambda$
\cite{rrr} since it is only their ratio which is fixed around $10^3$
TeV: the hierarchy $\sqrt{ F}, M\gg\Lambda$ is certainly allowed
\cite{raby}.  

The secluded sector often has accidental symmetries analogous 
to the baryon number. This means that the lightest particle
in the secluded sector might be stable and a good candidate for dark
matter with a mass of the order of $\sqrt{F}$, much larger than the
weak scale. The lightest messenger field might also be a good
candidate for superheavy DM. Indeed, if the supersymmetry breaking
sector contains only singlets under the $SU(3)_C\otimes SU(2)_L\otimes
U(1)_Y$ gauge interactions and if there are no direct couplings
between the ordinary and messenger sectors, then the theory is
characterized by a conserved global quantum number carried only by the
messenger fields. The typical mass $M$ of the DM component in the
messenger sector may be much larger than the weak scale.

Another framework in which we might expect the presence of superheavy
stable particles is a Kaluza-Klein theory (a unified theory which
requires space-time dimensions higher than four). A popular example is
provided by M-theory \cite{mtheory} where the number of dimensions is
$D=11$.  These theories are characterized by the presence of a tower
of Kaluza-Klein modes which are left after the compactification of the
extra $D-4$ dimensions. For instance, if $D=5$, the existence of a
compact fifth dimension implies an infinite tower of four-dimensional
particles corresponding to quantized excitations of the extra
dimension. These massive particles have been called ``pyrgons''
\cite{pyrgon}. If any of the pyrgon states are stable or have a
lifetime greater than the age of the universe, they might act as DM
with a mass of the order of the inverse of the physical size of the
compact dimensions $R_D^{-1}$, which is likely to be larger than the
weak scale by many orders of magnitude.

For the gravitational production scenario to be distinguishable from
other scenarios, $X$ must never thermalize.  The condition for the
dark matter particles to be out of equilibrium and their comoving
number density to be constant is
\begin{equation}
n_{X} \langle \sigma_A |v| \rangle \la H,  
\label{eq:equilib2}
\end{equation}
where $H$ is the Hubble parameter, $\langle \sigma_A |v| \rangle$ is
the thermal averaged self-annihilation cross section times the
M{\o}ller speed for the dark matter particles $X$.  Since the cross
section $\sigma_A$ is expected to be at most about $M_X^{-2}$ (usually
smaller; sometimes much smaller\footnote{For example, if there is a
heavy gauge particle mediating the process, then the effective
coupling will be further suppressed and the relevant mass scale for
the cross section will be the mediating particle mass instead of the
$X$ mass.}) and $n_X$ is bounded by the condition that $\Omega_X h^2
< 1$, we obtain from \eqr{eq:omegachi}
\begin{equation}
\frac{n_{X} \langle \sigma_A |v| \rangle}{H} \approx \frac{7 \times
10^{-19}}{ (T_{RH}/10^9 \mbox{GeV})} \
\frac{(H/\mpl)}{(M_X/\mpl)^3}
\end{equation}
as the quantity which must be less than one at $t=t_e$ to avoid
thermalization. For a low reheating temperature of $10^2$ GeV and a
typical value of $H=10^{-6} \mpl$ for inflationary scenarios, we find
a conservative condition $M_X/H \ga 1$ for the particles never to
reach chemical thermal equilibrium.  Note that this is a rather
conservative estimate since the reheating temperature is likely to be
larger and the cross section is likely to be smaller.  We also remark
that because the reheating temperature is likely to be much smaller
than the $X$ mass, the thermal production of the $X$ particles is
negligible.\footnote{Since for times larger than $t_e$, the
interaction rate continues to be smaller than $H$, the particles will
not thermalize later either.}

Now let us describe the basic physics underlying our mechanism of
gravitational production of DM.

In this paper we take space-time both in and out of inflationary era
to be spatially flat, homogeneous, and isotropic, with the line
element of the form
\begin{equation}
ds^2=a^2(\eta) (d\eta^2 - d{\bf x}^2).
\end{equation}
For simplicity (and without much loss of generality), we restrict
ourselves to a massive scalar field coupled to classical gravity and
nothing else.  The other couplings are assumed to play an
insignificant role in the gravitational production.

There are various inequivalent ways of calculating the particle
production due to interaction of a classical gravitational field with
the vacuum (see for example \cite{fulling}, \cite{birrelldavies}, and
\cite{chitre}).  In our work, we use the method of finding the
Bogoliubov coefficient for the transformation between positive
frequency modes defined at two different times.  We will show below
that the large mass dependence of the DM number density is determined
by either the differentiability (or the smoothness) of the scale
factor or the choice of the vacuum.  On the other hand, for $M_X/H \la
1$ where $H$ is the value at the end of inflation, the results are
quite insensitive to the differentiability or the fine details of the
scale factor's time dependence.  For $0.04 \la M_X/H \la 2$, we find
that all the dark matter needed for closure of the universe can be
made gravitationally, quite independently of the details of the
transition between the inflationary phase and the matter dominated
phase.

To see the effects of vacuum choice and the scale factor
differentiability on the large $X$ mass behavior of the $X$ density
produced, we start with the canonical quantization of the $X$ field in
an action of the form (in the coordinate $ds^2= dt^2- a^2(t)
d\bf{x}^2$)
\begin{equation}
S=\int dt \int d^3\!x\, \frac{a^3}{2}\left( \dot{X}^2 - \frac{(\nabla
X)^2}{a^2} - M_X^2 X^2 - \xi R X^2 \right)
\end{equation}
where $R$ is the Ricci scalar.  After transforming to conformal time
coordinate, we use the mode expansion
\begin{equation}
X({\bf x})=\int \frac{d^3\!k}{(2 \pi)^{3/2} a(\eta)} \left[a_k h_k(\eta) e^{i
{\bf{k \cdot x}}} + a_k^\dagger h_k^*(\eta) e^{-i {\bf{k \cdot x}}}\right],
\end{equation}
where because the creation and annihilation operators obey the
commutator $[a_{k_1}, a_{k_2}^\dagger] = \delta^{(3)}({\bf k}_1 -{\bf
k}_2)$, the $h_k$s obey a normalization condition $h_k h_k^{'*} - h_k'
h_k^* = i$ to satisfy the canonical field commutators (henceforth,
all primes on functions of $\eta$ refer to derivatives with respect to
$\eta$).  The resulting mode equation is
\begin{equation}
h_k''(\eta) + w_k^2(\eta) h_k(\eta) = 0,
\label{eq:modeequation}
\end{equation}
where 
\begin{equation}
w_k^2= k^2 + M_X^2 a^2 + (6 \xi - 1) a''/a \ .
\label{eq:frequency}
\end{equation}
The parameter $\xi$ is 1/6 for conformal coupling and 0 for minimal
coupling.  From now on, we will set $\xi=1/6$ for simplicity but
without much loss of generality.  By a change in variable $\eta
\rightarrow k/a$, one can rewrite the differential equation such that
it depends only on $H(\eta)$, $H'(\eta)/k$, $k/a(\eta)$, and
$M_X$.\footnote{This differential equation is $h_k''(y) + (1/H)H'(y)
h_k'(y) + (1+ M_X^2/y^2)/H^2\!(y) h_k =0$, where $y=k/a$.}  Hence, we
introduce the parameter $H_i$ and $a_i$ corresponding to the Hubble
parameter and the scale factor evaluated at an arbitrary conformal
time $\eta_i$, which we take to be the approximate time at which $X$
are produced (i.e., $\eta_i= \eta(t_e)$).  We then rewrite
\eqr{eq:modeequation} as 
\begin{equation}
h_\ktilde''(\etatilde) + (\ktilde^2 + b^2 \tilde{a}^2) h_\ktilde(\etatilde) 
=0 \qquad [b\equiv M_X/H_i]
\label{eq:scaledmodeequation}
\end{equation}
where $\etatilde=\eta a_i H_i$, $\tilde{a}=a/a_i$, and $\ktilde=
k/(a_i H_i)$.  For simplicity of notation, we shall drop all the
tildes from now on.  This differential equation can be solved once the
boundary conditions are supplied.  Since the annihilation operator is
just a coefficient of an expansion in a particular basis, fixing the
boundary conditions is equivalent to fixing the vacuum.

To obtain the number density of the particles produced, we will
perform a Bogoliubov transformation from the vacuum mode solution with
the boundary condition at $\eta=\eta_0$ (the initial time at which the
vacuum of the universe is determined) into the one with the boundary
condition at $\eta= \eta_1$ (any later time at which the particles are
no longer being created).  In the examples given in the next section,
$\eta_0$ will be taken to be $-\infty$ while $\eta_1$ will be taken to
be at $+\infty$ in order to define vacua of infinite adiabatic order
(explained below) which results in a smaller particle production than
for any finite adiabatic order vacua.\footnote{In the numerical
calculation, one can only approximate these infinities with large
numbers, but the limit is not singular.} The exact values of $\eta_0$
and $\eta_1$ are not important for those examples as long as they are
in a region in which $a'/a^2 \ll 1$ or $b a/k \ll 1$.  Defining the
Bogoliubov transformation as $ 
h_k^{\eta_1}(\eta)= \alpha_k h_k^{\eta_0}(\eta) + \beta_k h_k^{*
\eta_0}(\eta)$ (the superscripts denote where the boundary condition
is set), we have the following energy density in the
particles produced: 
\begin{equation}
 \rho_X(\eta_1) = M_X n_X(\eta_1) = M_X
H_i^3\left (\frac{1}{\tilde{a}(\eta_1)}\right)^3 \int_0^{\infty}
\frac{d\tilde{k}}{2
\pi^2} \tilde{k}^2 |\beta_{\tilde{k}}|^2, 
\end{equation} 
where\footnote{Here we restored the tildes for clarity.} one should
note that the number operator is defined at $\eta_1$ while the quantum
state (approximated to be the vacuum state) defined at $\eta_0$ does
not change in time in the Heisenberg representation.

As usual, there is an ambiguity in the definition of the vacuum, which
is equivalent to an ambiguity in the boundary conditions of
\eqr{eq:modeequation}.  One method of systematically classifying the
various inequivalent vacuum states is through the adiabatic vacuum
\cite{bunch} definition.  The adiabatic vacuum definition allows one
to construct and classify a set of mode equation solutions which
reduce to the usual plane waves when $a'(\eta)=0$ for all $\eta$.  The
classification is based on a type of WKB asymptotic expansion in
powers of conformal time derivatives of $w_k$.  In particular, the
classification allows one 
to quantify how two solutions with different boundary conditions
(hence two vacua) will differ in terms of derivatives of $w_k$.  Each
derivative with respect to the conformal time is assigned a
bookkeeping small parameter, and this small parameter's power in an
expansion is referred to as the adiabatic order.  We define $A$th
adiabatic (order) vacuum at time $\eta^*$ by using the boundary
condition  
\be
h_k(\eta^*) = h_k^{(A)}(\eta^*), \qquad
h_k'(\eta^*) = h_k^{'(A) }(\eta^*),
\ee
where $h_k^{(A)}(\eta)$ is a systematically chosen approximate
solution to the mode equation that satisfies the mode equation up to
$A$th adiabatic order in the asymptotic limit that the adiabatic
parameter goes to zero.  Roughly speaking, the larger
the adiabatic order of the vacuum, the closer it is to the Minkowski
vacuum in the sense that it is less (in the adiabatic limit) dependent
on the time at which it is defined.  We refer the reader to the
appendix (or Ref.\ \cite{birrelldavies}) for a more precise
definition.   

As shown in the appendix, the asymptotic behavior of the number
density as $b \rightarrow \infty$ can be obtained by the following
rule: If the vacuum at $\eta_0$ corresponds to an $n$th adiabatic
vacuum, and the vacuum at $\eta_1$ corresponds to a $p$th adiabatic
vacuum, then as $b \rightarrow \infty$ the number density will behave
like\footnote{This behavior is also noted on pg.\ 69 of
\cite{birrelldavies} although there it is arrived at
differently than in our Appendix.}
\begin{equation}
n_X \sim b^{-(2 r+ 1)}
\label{eq:basympt}
\end{equation}
where $r= \mbox{Min}(p,n)$ provided that $(d^\nu a/ d\eta^\nu) /
a^{\nu+1} < \infty$ for all $\eta \in [\eta_0, \eta_1]$ and all
natural numbers $\nu$.\footnote{Note that by definition given in
the appendix, $p$ and $n$ can be only even natural numbers or 0.} It is
important to note that for a fixed time, the asymptotic
expansion generated by \eqr{eq:wkbiter} (in the Appendix)
generally only converges up to a finite order.  Hence, except
under special circumstances an adiabatic vacuum of only a finite
order can be generated.  This means that in general, the number
density will fall off with a finite power of $1/b$ for large
$b$.  Only when an adiabatic vacuum of infinite adiabatic order
can be generated, which usually means that the domain of
$a(\eta)$ can be extended to $\pm \infty$ with the property
given above, does the number of particles produced fall off
faster than any finite power of $1/b$ (e.g.,  exponential
suppression).  In practice, we find that a spacetime which
admits an infinite adiabatic order vacuum has the ``advantage''
of all the vacua defined in a sufficiently adiabatic region
being numerically equivalent regardless of the vacua's adiabatic
order and the exact time at which the vacua are defined.

If within the domain there is one discontinuity of the first
kind\footnote{The discontinuity of the first kind refers to the
situation where the left 
and the right hand limits exist but are unequal.} in $(d^q a/d
\eta^q)/a^{q+1}$ for $q=s$ where $-2 < s-2 \leq r=\mbox{Min}(p,n)$ and
there are no discontinuities for $q<s$,
\begin{equation}
n_X \sim  b^{-(2 s - 3)}.
\label{eq:simpdiscont}
\end{equation}
This is true provided that $(d^\nu
a/ d\eta^\nu) / a^{\nu+1} < \infty$ for all $\eta$ in each of the
continuous domain and all natural numbers $\nu$.  Note that fractional
power dependence on $1/b$ will be possible if the discontinuity is not
of the first kind (e.g., $a(\eta)= \eta^2$ at $\eta=0$).

%%%%%%%%%%%%%%%%%%%%%%%%%%%%%%%%%%%%%%%%
%%%%%%%%%%%%%%%%%%%%%%%%%%%%%%%%%%%%%%%%
\vspace{36pt}
\centerline{\bf III. NUMERICAL RESULTS}
\vspace{24pt}
%%%%%%%%%%%%%%%%%%%%%%%%%%%%%%%%%%%%%%%%
%%%%%%%%%%%%%%%%%%%%%%%%%%%%%%%%%%%%%%%%

We shall employ the method elaborated in the previous section to
calculate the gravitational production of particles in a couple of toy
models of inflation which reflect some extreme ranges of
differentiability.  We will consider the case when the scale factor
has a discontinuity of the first kind in one of its derivatives and
when it is a $C^\infty$ function.  We will see that enough dark matter
may be produced through this mechanism as to give critical density of
dark matter today.

Our example of a discontinuous model has the scale factor of the de
Sitter space for $\eta < \eta_i$ and the matter or radiation dominated
universe for $\eta > \eta_i$:
\begin{equation}
a(\eta) = \left\{ \begin{array}{ll} a_i/(2-\eta/\eta_i)
 & \qquad \mbox{if $\eta \leq \eta_i$}, \\ 
		a_i(\eta/\eta_i)^p  & \qquad \mbox{if $\eta > \eta_i$},
		    \end{array}
		\right.
\end{equation}
where $p=2$ for the matter dominated case and $p=1$ for the radiation
dominated case.  If we define our vacuum states at $\eta_0=-\infty$
and $\eta_1=\infty$, then in this space time, adiabatic vacua of any
order will be equivalent to infinite-order adiabatic vacua.  For the
transition into matter domination, the smallest derivative order in
which there is a discontinuity (of the first kind) comes from
$a'(\eta)$ at $\eta=\eta_i$ while for the transition into the
radiation domination, the analogous contribution comes from
$a''(\eta)$ which has a discontinuity of the first kind at $\eta=
\eta_i$.\footnote{These discontinuities are unphysical and correspond
at best to crude approximations.  We consider them to test the
sensitivity of our results on the choice of the model.} Hence, our
analysis would predict that for large $b$, $n_X$ will increase like
$b$ in the case of matter domination whereas it will fall off like
$1/b$ in the case of radiation domination.

%%%%%%%%%%%%%%%%%%%%%%%%%%%%%%%%%%%%%%%%
%%%%%%%%%%%%%%%%%%%%%%%%%%%%%%%%%%%%%%%%
\begin{figure}[t]
\hspace*{25pt} \epsfxsize=400pt \epsfbox{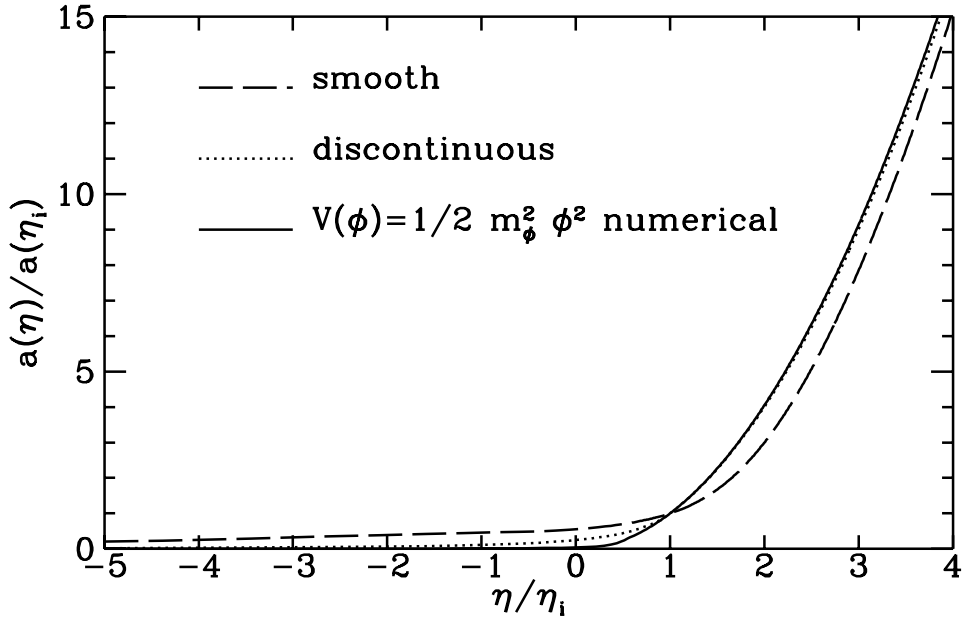}
\caption{The scale factor (normalized to its value at $\eta=\eta_i$)
is plotted as a function of the scaled conformal time $\eta/\eta_i$.
The curves labeled ``smooth'' and ``discontinuous'' correspond to the
toy models for the evolution of the scale factor without and with
discontinuous derivatives, respectively.  As $\eta/\eta_i \rightarrow
- \infty$, the scale factor behaves as $1/(\eta/\eta_i)^2$
corresponding to a de Sitter space, and as $\eta/\eta_i \rightarrow
+\infty$, the scale factor behaves as $(\eta/\eta_i)^2$ corresponding
to a matter dominated Universe.  The solid curve shows an analogous
numerical solution of the scale factor for a $(1/2) m_\phi^2 \phi^2$
inflaton potential.}
\label{fig:potentials}
\end{figure}
%%%%%%%%%%%%%%%%%%%%%%%%%%%%%%%%%%%%%%%%
%%%%%%%%%%%%%%%%%%%%%%%%%%%%%%%%%%%%%%%%

Our second toy model looks at the other extreme limit of having a $C^\infty$
function for $a(\eta)$ which behaves in the asymptotic limits of
$\eta/\eta_i \rightarrow \pm \infty$ identically as the discontinuous model:
\begin{eqnarray}
a( \eta ) &  =  & a_i \left\{
\frac{1-\exp[-(\eta/\eta_i)^2]}{(\eta/\eta_i)^2} \left(
\frac{1-\tanh(\eta/\eta_i/2+1)}{2} \right) + 
\frac{(\eta/\eta_i)^{2 p}}{(1 + 3 \exp(-\eta/\eta_i))^2} \right. \nonumber \\
& & \left. \phantom{\frac{A}{B}} 
+ \tanh(\eta/\eta_i- \lambda) - \tanh(\eta/\eta_i-2 \lambda) \right\}^{1/2} 
\end{eqnarray}
where $\lambda=1.07$ is needed for proper normalization.  The
functional form was chosen rather arbitrarily except for the
requirements of monotonicity, $(d^\nu a/d\eta^\nu)/a^{\nu+1} < \infty$
for all $\eta$ and natural numbers $\nu$, and appropriate power law
asymptotic behavior.  As before, this spacetime admits a vacuum of
infinite adiabatic order at $\eta=\pm \infty$.  In Fig.\
\ref{fig:potentials} we show how these models compare with the
numerical result obtained by solving the $(1/2) m_\phi^2 \phi^2$
inflationary model's equations of motion.  Note that they differ
mainly in the transition region (or the ``nonadiabatic'' region near
$\eta/\eta_i=1$) where most of the particle production ``occurs.''

%%%%%%%%%%%%%%%%%%%%%%%%%%%%%%%%%%%%%%%%
%%%%%%%%%%%%%%%%%%%%%%%%%%%%%%%%%%%%%%%%
\begin{figure}[t]
\hspace*{25pt} \epsfxsize=400pt \epsfbox{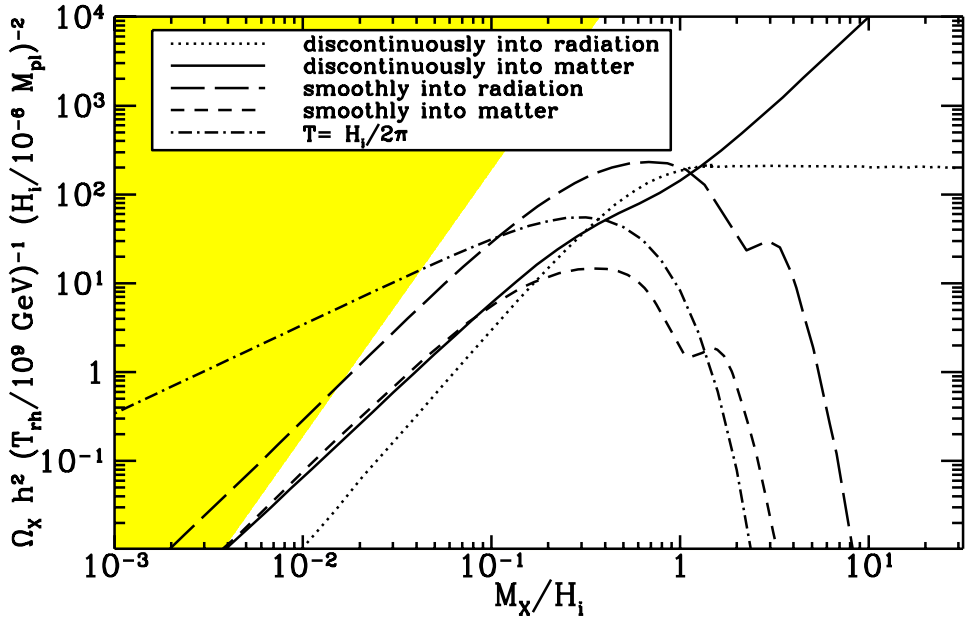}\\
\caption{ The dark matter abundance today is shown as a function of
the particle mass for various models.  The mass is given in terms of
$H_i \approx 10^{-6} \mpl$ (the Hubble parameter at $\eta=\eta_i$, the
beginning of the coherent oscillation period).  In the
``discontinuously into radiation'' case, $a''(\eta)$ has a
discontinuity at $\eta=\eta_i$, while in the ``discontinuously into
matter'' case, $a'(\eta)$ has a discontinuity at $\eta=\eta_i$.  The
curves labeled ``smoothly into'' is for $a(\eta)$ that satisfies
$(d^\nu a/d \eta^\nu)/a^{\nu+1} < \infty$ for all $\eta$ and natural
numbers $\nu$.  The curve labeled $T=H_i/(2 \pi)$ shows a thermal
density with this temperature.  The unshaded region satisfies the {\it
conservative} nonthermalization condition obtained by setting $\langle
\sigma_A \protect{|}v\protect{|} \rangle = 1/M_X^2$ in Eq.\
(\protect{\ref{eq:equilib2}}).} 
\label{fig:mainresult}
\end{figure}
%%%%%%%%%%%%%%%%%%%%%%%%%%%%%%%%%%%%%%%%
%%%%%%%%%%%%%%%%%%%%%%%%%%%%%%%%%%%%%%%%

In Fig.\ \ref{fig:mainresult}, we show the number density obtained
numerically in these toy models.  The peak at $M_X/H_i \sim 1$ for the
$C^\infty$ model is similar to the case presented in Ref.\
\cite{birrelldavies1980}.  As we expect, the large $b$ behavior
is exactly determined by the choice of the vacuum and the
differentiability of the potential.  Specifically, just as our
asymptotic analysis showed, for large $b$, $n_X$ varies as $b$ and
$1/b$ for the matter and radiation domination case respectively in the
discontinuous model, while $n_X$ is exponentially suppressed in the
$C^\infty$ model.  Note that independently of the differentiability of
the scale factor, if $M_X \approx H_i \approx m_\phi$ for
$T_{RH}\approx 10^9$ GeV, $X$ will have critical energy density today.
On the other hand, unless there is some discontinuity in the scale
factor for some $n$th derivative where $n$ is not too large, this
gravitational production mechanism will not generate enough dark
matter in the Universe to give critical density for much larger masses
($M_X \gg m_\phi$) even if such stable heavy particles exist.
Furthermore, even in the mass range in which the density of particles
produced peaks, if the reheating temperature is below about $10^7$
GeV, this mechanism will most likely not generate significant amount
of dark matter.  In the case that this mechanism cannot produces these
heavy particles, however, if these heavy particles couple to the
inflaton and if preheating occurs, enough of them may be produced
through the broad resonance mechanism to have critical density of
superheavy dark matter today.

%%%%%%%%%%%%%%%%%%%%%%%%%%%%%%%%%%%%%%%%
%%%%%%%%%%%%%%%%%%%%%%%%%%%%%%%%%%%%%%%%
\vspace{36pt}
\centerline{\bf IV. SUMMARY}
\vspace{24pt}
%%%%%%%%%%%%%%%%%%%%%%%%%%%%%%%%%%%%%%%%
%%%%%%%%%%%%%%%%%%%%%%%%%%%%%%%%%%%%%%%%

To conclude, we have investigated the scenario of creating
nonthermalizing dark matter gravitationally at the end of inflation
(or the beginning of the coherent oscillation phase).  There is a
significant mass range ( $0.1m_\phi$ to $m_\phi$, where $m_\phi
\approx 10^{13}$GeV) for which the $X$ particles will have critical density
today regardless of the fine details of the inflation-matter/radiation
transition.  Because this production mechanism is inherent in the
dynamics between the classical gravitational field and a quantum
field, it needs no fine tuning of field couplings or any coupling to
the inflaton field.  However, only if the particles are stable (or
sufficiently long lived) will these particles give contribution of the
order of critical density.  For even larger dark matter masses, the
broad resonance mechanism of preheating (if it occurs) will produce
these particles in sufficient abundance as to achieve $\Omega_0 = 1$.

%%%%%%%%%%%%%%%%%%%%%%%%%%%%%%%%%%%%%%%%
%%%%%%%%%%%%%%%%%%%%%%%%%%%%%%%%%%%%%%%%
\vspace{36pt}
\centerline{\bf ACKNOWLEDGMENTS}
\vspace{24pt}
%%%%%%%%%%%%%%%%%%%%%%%%%%%%%%%%%%%%%%%%%%%%%
%%%%%%%%%%%%%%%%%%%%%%%%%%%%%%%%%%%%%%%%%%%%%
We would like to thank Robert Wald, Ewan Stewart, Andrei Linde, Lev
Kofman, Norbert Sch\"{o}rghofer, Alvaro de Rujula, and Christoph Schmid for
discussions.  We would like to also thank the referee for some helpful
suggestions.   DJHC and EWK were supported by the DOE and NASA under
Grant NAG5-2788.

%%%%%%%%%%%%%%%%%%%%%%%%%%%%%%%%%%%%%%%%
%%%%%%%%%%%%%%%%%%%%%%%%%%%%%%%%%%%%%%%%
\frenchspacing
\def\prpts#1#2#3{Phys. Reports {\bf #1}, #2 (#3)}
\def\prl#1#2#3{Phys. Rev. Lett. {\bf #1}, #2 (#3)}
\def\prd#1#2#3{Phys. Rev. D {\bf #1}, #2 (#3)}
\def\plb#1#2#3{Phys. Lett. {\bf #1B}, #2 (#3)}
\def\npb#1#2#3{Nucl. Phys. {\bf B#1}, #2 (#3)}
\def\apj#1#2#3{Astrophys. J. {\bf #1}, #2 (#3)}
\def\apjl#1#2#3{Astrophys. J. Lett. {\bf #1}, #2 (#3)}
%%%%%%%%%%%%%%%%%%%%%%%%%%%%%%%%%%%%%%%%
%%%%%%%%%%%%%%%%%%%%%%%%%%%%%%%%%%%%%%%%
\begin{picture}(400,50)(0,0)
\put (50,0){\line(350,0){300}}
\end{picture}

%%%%%%%%%%%%%%%%%%%%%%%%%%%%%%%%%%%%%%%%
%%%%%%%%%%%%%%%%%%%%%%%%%%%%%%%%%%%%%%%%
\vspace{36pt}
\centerline{\bf APPENDIX}
\vspace{24pt}
%%%%%%%%%%%%%%%%%%%%%%%%%%%%%%%%%%%%%%%%
%%%%%%%%%%%%%%%%%%%%%%%%%%%%%%%%%%%%%%%%
\setcounter{equation}{0}
\renewcommand{\theequation}{A\arabic{equation}}

In this appendix, we derive \eqr{eq:basympt} and
\eqr{eq:simpdiscont}, the asymptotic dependence of the dark
matter density on the mass parameter $b=M_X/H_i$ as $b
\rightarrow \infty$.  This asymptotic behavior is, in general,
dependent upon the choice of the vacuum state and the
differentiability of the scale factor in an FRW type spacetime.
We employ the adiabatic vacua ansatz \cite{bunch} to classify the
various possible (restricted) choices of vacua.  Strictly
speaking, our conditions for the various asymptotic behaviors
are only sufficient conditions, but they have wide applicability
as we demonstrate in this paper.

Let us first review the concept of an adiabatic vacuum (see for
example pg.\ 66 of Ref.\ \cite{birrelldavies}).  We first define the
concept of an adiabatic order as the power of $1/T$ that results
for any term in a $1/T$ expansion after one makes the
transformation $\eta \rightarrow \eta$ and $d/d\eta \rightarrow
T^{-1} d/d\eta$.  Note that if $T^{-1} \rightarrow 0$, then this is
equivalent to an expansion in ``smallness'' of conformal time
derivatives.  The basic idea is that if the derivatives of the
mode frequency $w_k$ are indeed small, then the degree to which
the field theory breaks time translational symmetry can be
characterized by the adiabatic order.  This breaking of the time
translational symmetry\footnote{Conformal time translation generates
conformal transformation in an FRW universe, and the mass term
breaks the conformal symmetry.}  is what is responsible for
particle creation in our isotropic expanding Universe.

To define the adiabatic vacuum, we first make a change in
variables from $h_k$ to $W_k$ by writing \be
h_k=\frac{1}{\sqrt{2 W_k}} \exp\left(-i \int^\eta W_k(\eta')
d\eta'\right)
\label{eq:newsol}
\end{equation}
and obtain a new differential equation
\be
W_k^2 = w_k^2 - 1/2 \left[W_k''/W_k - (3/2) (W_k'/W_k)^2\right],
\label{eq:wkbiter}
\end{equation}
where we have used \eqr{eq:scaledmodeequation} and defined $w_k^2$ to be
the coefficient of $h_\ktilde$ in
\eqr{eq:scaledmodeequation}.\footnote{We have dropped all the tildes
for simplicity in notation.  Note also that a constant factor
normalization choice of \eqr{eq:newsol} is unimportant for the
Bogoliubov transformation.} 
Hence, let us define a map
\be
A[W_k^{(n)}]= \sqrt{ w_k^2 - \frac{1}{2} \left[ \frac{W_k^{''(n)}}{W_k^{(n)}} - \frac{3}{2}
\left(\frac{W_k^{'(n)}}{W_k^{(n)}}\right)^2\right] }
\label{eq:coolmap}
\end{equation}
which is a map that raises the adiabatic order by two and also 
define 
\be
W_k^{(n+2)}= A[W_k^{(n)}],
\label{eq:recursion}
\end{equation}
where the superscript denotes the adiabatic order and $W_k^{(0)}=
w_k$.  We can now write an approximate mode equation
solution\footnote{After finishing our paper, 
we learned that a complete and more precise analysis of the asymptotic
behavior of 
the adiabatic modes  can be found in Ref. \cite{olver}.}
good to $A$th adiabatic order as
\be
h_k^{(A)} = \frac{1}{\sqrt{2 W_k^{(A)}}} \exp\left(-i \int^\eta
W_k^{(A)}(\eta') d\eta'\right)\ .
\end{equation}
Finally, we define the adiabatic vacuum of $A$th order at some value
of $\eta$ which we call $\eta^*$ by using the boundary
condition\footnote{In the spirit of the adiabatic expansion, the
equality needs to only be enforced to $A$th adiabatic order.  However,
we will for simplicity of argument assume throughout that it is
enforced exactly.}
\be h_k(\eta^*) = h_k^{(A)}(\eta^*), \qquad h_k'(\eta^*) = h_k^{'(A)
}(\eta^*),
\label{eq:adboundcond}
\end{equation}
where $h_k$ on the left hand side solves the mode equation
\eqr{eq:scaledmodeequation} exactly.  Since for a generic finite
$\eta^*$ and fixed $b$, the recursion generated by
\eqr{eq:recursion} eventually increases without bound in
general, the recursion relation generates at best an asymptotic
expansion in the limit that the higher than zeroth adiabatic order
terms go to zero.  In particular, an infinite adiabatic order vacuum
usually cannot be generated at a ``nonsingular'' $\eta^*$.

Now, we examine how different boundary conditions (different adiabatic
order vacua) give rise to different asymptotic behaviors as $b
\rightarrow \infty$.  First let us restrict our attention to the case
where $a(\eta)$ is $C^\infty$ in the domain of interest.  Since $b^2$
is the coefficient of $a^2$ term inside $w_k^2$, we see that a
sufficient condition for the higher than zeroth adiabatic order terms to
go to zero for large $b$ is $(d^\nu a/ d\eta^\nu) / a^{\nu+1} <
\infty$ for all $\eta$ in the domain and any finite natural number
$\nu$.  Hence, we will assume this to be true and use the adiabatic
expansion to determine the asymptotic power dependence of $n_X$ as
$1/b \rightarrow 0$.

The key is that the recursion \eqr{eq:recursion} can be used as a
generator of an asymptotic expansion of the exact solution in the
limit that the higher than 0th adiabatic order terms tend to zero.
One can easily show that this map has the property if
$A[W_k^{(n)}]/W_k^{(n)} \sim 1 + {\cal O}(1/b^\alpha) + {\cal
O}(1/b^{\alpha + \mu})$ with $\mu \geq 1$, then
$A[A[W_k^{(n)}]]/A[W_k^{(n)}] \sim 1 + {\cal O}(1/b^{\alpha + 2}) +
{\cal O}(1/b^{\alpha + 2 + \mu})$ where $\sim$ represents the
asymptotic limit that $b \rightarrow \infty$.  Since $W_k^{(2)} = w_k
+ {\cal O}(1/b)$, we arrive at an useful property \be W_k^{(n)}=
W_k^{(n-2)} + {\cal O}(1/b^{n-1}),
\label{eq:useful}
\end{equation}
which shows how each successive approximation generates corrections of
only increasingly higher order in $1/b$.  Thus,
\be 
W_k(\eta) \sim w_k(\eta) + \sum_{n=0}^{A/2-1}
(W_k^{(2 n+2)} - W_k^{(2 n)}) + {\cal O}(1/b^{A+1})
\ee
is an asymptotic expansion of the solution to \eqr{eq:wkbiter} with
the boundary condition
\be
W_k(\eta^*)=W_k^{(A)}(\eta^*) + h(\eta^*), \qquad W_k'(\eta^*)=W_k^{'(A)}(\eta^*) + h'(\eta^*)
\label{eq:twidbc}
\ee
where $h(\eta) \sim {\cal O}(1/b^{A+1})$.  Let us call this solution
$f_k$.  Note that $f_k$ satisfies a boundary condition that differs
from one implied by \eqr{eq:adboundcond} by ${\cal O}(1/b^{A+1})$.

To check this is the asymptotic expansion for the solution satisfying
a different boundary condition (i.e. the one specified by
\eqr{eq:adboundcond}), one can now perturb about $f_k$ by writing
$W_k^{\eta^*}(\eta)=f_k(\eta) + u_k(\eta)$ where the superscript on
$W_k^{\eta^*}$ corresponds to the time at which the boundary condition
\eqr{eq:adboundcond} is imposed and by using this in
\eqr{eq:wkbiter} to obtain a differential 
equation linear in $u_k$.  One then finds that the sourced solution
contributes only ${\cal O}(1/b^{A+1})$ to $u_k(\eta)$.  Hence, if
the initial data on $u_k(\eta)$ is of the order of ${\cal
O}(1/b^{A+1})$, then the behavior of $u_k(\eta)$ as $b \rightarrow 
\infty$ is ${\cal O}(1/b^{A+1})$.  In particular,
if we have the boundary condition \eqr{eq:adboundcond} instead of
\eqr{eq:twidbc}, then we find $u_k(\eta^*) \sim {\cal O}(1/b^{A+1})$
and
\be
W_k^{\eta^*}(\eta) \sim W_k^{(A)}(\eta) + {\cal O}(1/b^{A+1})
\label{eq:behave}
\ee
where the ${\cal O}(1/b^{A+1})$ vanishes at $\eta=\eta^*$.  This is
of course what we would naively expect. 

We can now see how $n_X$ will depend asymptotically on $b$.  Suppose
the vacuum in the past is defined at $\eta=\eta_0$ with $n$th
adiabatic order boundary condition and the vacuum today is defined at
$\eta=\eta_1$ with $p$th adiabatic order boundary condition.  Carrying
out the Bogoliubov transformation with the solution written in the
form \eqr{eq:newsol}, we find
\be
| \beta_k(\eta_1, \eta_0)|^2 = \frac{1}{4 W_k^{\eta_0} W_k^{\eta_1}}
\left\{ \frac{1}{4} \left( \frac{W_k^{' \eta_0}}{W_k^{\eta_0}} -
\frac{W_k^{'\eta_1}}{W_k^{\eta_1}} \right)^2 + ( W_k^{\eta_0}-
W_k^{\eta_1})^2 \right\},
\label{eq:numdensasymp}
\end{equation}
where the right hand side can be evaluated at any $\eta$.  In light of
\eqr{eq:behave}, if we substitute $W_k^{\eta_0} = W_k^{(n) } + {\cal
O}(1/b^{n+1})$ and $W_k^{\eta_1} = W_k^{(p)} + {\cal O}(1/b^{p+1})$, then
$W_k^{\eta_0} - W_k^{\eta_1}= {\cal O}(1/ b^{r+1})$ where $r=
\mbox{Min}(n,p)$ and $W_k^{' \eta_0 }/W_k^{\eta_0} - W_k^{' \eta_1
}/W_k^{\eta_1} = {\cal O}(1/b^{r+2})$.  Now, since
\be
n_X \propto \int_0^\infty |\beta_k|^2 k^2 dk,
\end{equation}
after making a change of variable from $k$ to $y$ through
$k = y b$, we obtain the result in \eqr{eq:basympt}.  

If within the domain of interest there is one discontinuity of the
first kind (left and right hand limits exist but are unequal) in $(d^q
a/d \eta^q)/a^{q+1}$ at $\eta=\eta_d$ for $q=s$ where $-2 < s-2 \leq
r=\mbox{Min}(p,n)$, and there are no discontinuities for $q<s$, then
\eqr{eq:numdensasymp} will receive leading contributions at the
discontinuity.  Note that the asymptotic expansion is valid in each
``continuous'' region because the discontinuity is of the first kind.
Hence, with similar considerations as with the smooth case above, we
can obtain \eqr{eq:simpdiscont}.  However, unlike in the continuous
case, the asymptotic expansion can be used to evaluate
\eqr{eq:numdensasymp} only at $\eta=\eta_d$ because the asymptotic
expansion cannot be extended beyond each of the continuous regions.
If $s$ is even, then $W_k^{\eta_0}(\eta_d)-W_k^{\eta_1}(\eta_d) \sim
{\cal O}(1/b^{s-1})$ will give the leading contribution in
\eqr{eq:numdensasymp}  because $W^{''(s-2)}/W^{(s-2)}$ is discontinuous.  If
$s$ is odd, then the leading contribution to \eqr{eq:numdensasymp}
will come from the difference
$W_k^{'\eta_0}(\eta_d)/W_k^{\eta_0}(\eta_d) - W_k^{' \eta_1}(\eta_d)
/W_k^{\eta_1}(\eta_d) \sim {\cal O}(1/b^{s-1})$ because
$(W^{''(s-3)}/W^{(s-3)})'$ is discontinuous.
\end{document}